\documentclass[aps,prb,twocolumn,groupedaddress,showpacs]{revtex4}

\usepackage{amssymb}
\usepackage{graphicx}
\usepackage{amsmath}
\usepackage{hyperref}

\begin{document}

\title{Josephson effect through an anisotropic magnetic molecule}

\author{
	I.~A.~Sadovskyy,$^{1}$
	D.~Chevallier,$^{2,3}$
	T.~Jonckheere,$^{2}$
	M.~Lee,$^{4}$
	S.~Kawabata,$^{5}$, and
	T.~Martin$^{2,3}$
}

\affiliation{
	$^{1}$Rutgers University,
	136 Frelinghuysen Road, Piscataway, New Jersey, 08854, USA
}

\affiliation{
	$^{2}$Centre de Physique Th\'eorique, 
	Case 907 Luminy, F-13288 Marseille Cedex 9, France
}

\affiliation{
	$^{3}$Universit\'e de la M\'edit\'erann\'ee, 
	F-13288 Marseille Cedex 9, France
}

\affiliation{
	$^{4}$Department of Applied Physics, College of Applied Science, Kyung Hee University, Yongin 446-701, Korea
}

\affiliation{
	$^{5}$Nanosystem Research Institute (NRI), National Institute of Advanced Industrial Science\\
	and Technology (AIST), and JST-CREST, Tsukuba, Ibaraki 305-8568, Japan
}

\date{\today}

\begin{abstract}
We study the Josephson effect through a magnetic molecule with anisotropic properties. Performing calculations in the tunneling regime, we show that the exchange coupling between the electron spin on the molecule and the molecular spin can trigger a transition from the $\pi$ state to the $0$ state, and we study how the spin anisotropy affects this transition. We show that the behavior of the critical current as a function of an external magnetic field can give access to valuable information about the spin anisotropy of the molecule.
\end{abstract}


\maketitle

\section{Introduction}
\label{sec:introduction}

The Josephson effect\cite{Josephson:1962,Gennes:1964} is a striking manifestation of many body physics and macroscopic quantum coherence in condensed matter systems. While early investigations concerned mainly bulk superconducting junctions separated by an insulating barrier, in the last decades it has become a very active field of study in the context of mesoscopic physics. Indeed the insulating barrier can be replaced by a conductor or a nano-device that can be as small as a quantum dot or a single molecule. In this sense the study of the Josephson current can provide a novel way to investigate the electronic properties of the nano-object, which is sandwiched between the superconducting electrodes. More than a decade ago, it was predicted using the Krein theorem\cite{Spivak:1991,Krichevsky:2000,Benjamin:2007} that when a singly occupied quantum dot in the Coulomb blockade regime is inserted between the superconductors, the Josephson current phase relation acquires a $\pi$ shift, i.e., the critical current has the opposite sign from that of a tunnel junction. A phase diagram of the $\pi$--$0$ phase transition was derived later on for contacts with arbitrary transparency using a combination of Hubbard-Stratonovich and saddle-point approximation.\cite{Rozhkov:2000} Experimentally (for nanoscale devices) it was measured in superconductor--nanotube--superconductor systems.\cite{Dam:2006} This picture gets more complicated when the Kondo temperature is lower than the superconducting gap: a $0$-junction state is restored,\cite{Siano:2004,Choi:2004} albeit with a different current phase relationship.

In recent years theoretical and experimental studies have addressed transport geometries where a molecule~--- artificial or otherwise~--- is inserted between two electrodes.\cite{Kasumov:1999,JarilloHerrero:2006,Cleuziou:2006,Eichler:2009,Roch:2008,Winkelmann:2009} This goes one step beyond the study of transport through quantum dots because the molecule has internal degrees of freedom (such as vibrations and possibly spin). On the one hand, such degrees of freedom have an effect on the electronic current, on the other hand, the current itself can be considered as a probe of the inherent mechanisms of the molecule. 

A subfield of molecular electronics is called molecular spintronics: it focuses on molecules which have an intrinsic spin,\cite{Sessoli:2003,Petukhov:2004} and it is expected that electron transfer through the molecule can trigger changes in the molecule spin because of the existence of an exchange coupling with the electron spin. Such molecules (such as a buckminsterfullerene doped with a magnetic atom) may have an isotropic spin, or otherwise the spin may have a preferred direction due to the crystalline structure of the molecule (this is the case of Mn$_{12}$ acetate). Recently, there have been some efforts to describe and measure transport through molecular spintronics devices with normal metal or ferromagnetic leads,\cite{Romeike:2006,Heersche:2006a,Grose:2008,Roch:2011} with an emphasis on master equations approach on the theoretical side. Nevertheless, efforts in the field of molecular spintronics with superconducting electrodes are still at their beginning stage.

\begin{figure}[b]
	\includegraphics[width=4.5cm]{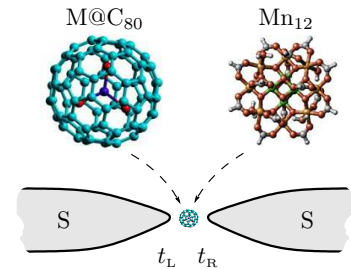}
	\caption{
(Color online) A magnetic molecule (e.g., M@C$_{80}$, Mn$_{12}$, $\ldots$) connecting two superconductors via tunnel barriers $t_{\rm\scriptscriptstyle L}$ and $t_{\rm\scriptscriptstyle R}$. The exchange coupling between the molecular spin and the electronic spin can strongly modify the Josephson current.
	}
	\label{fig:setup}
\end{figure}

A recent theoretical work focused on the Josephson current through an isotropic magnetic molecule, via perturbative calculations in the tunneling Hamiltonian as well as numerical renormalization group calculations.\cite{Lee:2008a} It allowed us to draw a complete phase diagram of the $\pi$--$0$ phase transition. An equivalent study of supercurrent through molecules which have an anisotropic spin, which magnetization can tunnel, and which are subject to a (weak) external magnetic field is still lacking. This is precisely the focus of the present work. One of the challenges of this work is that we have to deal with a large number of parameters: the exchange coupling $J$ between the dot electron spin and the molecule spin, the anisotropy constant $D$, and the coefficient $B_2$ for quantum tunneling of magnetization of the molecule, the dependence on external magnetic field $B$ as well as the dot level $\epsilon_{\rm d}$, which can be adjusted by a gate voltage. Note that it is now experimentally possible to manipulate the anisotropy parameters of magnetic molecules.\cite{Parks:2010,Zyazin:2010} One of our goals is to determine to what extent the measurement of the critical current can provide information of the sign or magnitude of such parameters. For simplicity, we focus on the regime where the superconducting gap is much larger than the Kondo temperature, which allows us to focus on weak coupling (small tunneling Hamiltonian) calculations. Also, we restrict the analysis on the simplest case of a molecule spin $S=1$ to demonstrate the effect where the two main contributions due to spin anisotropy (easy axis anisotropy and quantum tunneling of magnetization) are present.

The outline of the paper is as follows. In Sec.~\ref{sec:model}, we introduce the model for the magnetic molecule connected to two superconducting leads and we compute the expression of the Josephson current through this molecule. In Sec.~\ref{sec:results}, we study the effect of the anisotropic parameters and of the adjustable experimental parameters on the sign of the critical current. Finally, we conclude in Sec.~\ref{sec:conclusion}.

\section{Model}
\label{sec:model}

\subsection{Hamiltonian}

\begin{figure}[t]
	\includegraphics[width=4.2cm]{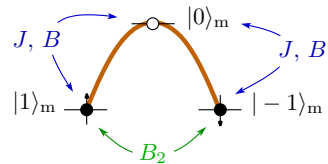}
	\caption{
(Color online) Spin states for $S=1$ spin, and coupling between these states due to the different terms of the Hamiltonian. $B_2$ induces tunneling between $|1\rangle_{\rm m}$ and $|-1\rangle_{\rm m}$ states; $J$ and $B$ induce $|1\rangle_{\rm m} \leftrightarrow |0\rangle_{\rm m}$ and $|0\rangle_{\rm m} \leftrightarrow |-1\rangle_{\rm m}$ tunnelings.
	}
	\label{fig:spin_states}
\end{figure}

The total Hamiltonian of the system (see Fig.~\ref{fig:setup}) consists of the three terms $\mathcal{\hat H} = \mathcal{\hat H}_{\rm d} + \mathcal{\hat H}_{\rm s} + \mathcal{\hat H}_{\rm t}$. The first one is the Hamiltonian of the molecule,
\begin{equation}
	\mathcal{\hat H}_{\rm d} = 
	\mathcal{\hat H}_{\rm m} 
	+ \epsilon_{\rm d} \sum\limits_\sigma {\hat d}_\sigma^\dag {\hat d}_\sigma^{\phantom\dag}
	+ U {\hat d}_\uparrow^\dag {\hat d}_\uparrow^{\phantom\dag} {\hat d}_\downarrow^\dag {\hat d}_\downarrow^{\phantom\dag},
	\label{eq:Hd}
\end{equation}
where $\epsilon_{\rm d}$ is the electronic level of the molecule implied in the transport, and $U$ is the Coulomb interaction strength. The ${\hat d}_\sigma^\dag$ and ${\hat d}_\sigma^{\phantom\dag}$ are electronic creation and annihilation operators on the electronic level in the molecule. Since~$U$ is typically much larger than the other energies in the system, we consider the limit of infinite Coulomb interaction $U\rightarrow\infty$, thus only one electron is allowed to occupy the dot. With this assumption, the Hamiltonian $\mathcal{\hat H}_{\rm m}$, which characterizes the magnetic properties of the dot, reads
\begin{equation}
	\mathcal{\hat H}_{\rm m} = 
	-D {\hat S}_z^2 + B ({\hat S}_z + {\hat s}_z)
	- \frac{B_2}{2} ({\hat S}_+^2 + {\hat S}_-^2) +
	J {\hat{\bf S}}{\hat{\bf s}},
	\label{eq:Hm}
\end{equation}
where $S_z$ is the molecular spin and $s_z$ the spin of the electron on the molecule (if present). $J$ is the exchange coupling between molecular and electronic spin, $D>0$ is the easy axis anisotropy constant, $B_2$ is the coefficient of quantum tunneling of magnetization (QTM), and $B$ is the external magnetic field. Figure~\ref{fig:spin_states} shows how these terms couple the states of the molecule in the case of a spin $S=1$. In order to avoid a too large number of parameters, we have made some simplifying assumptions when writing this Hamiltonian: the anisotropy terms are not affected by the charge of the dot level (this should be the case for systems like M@C$_{80}$, but not for molecules like Mn$_{12}$),\cite{Heersche:2006b} the magnetic field is taken parallel to the spin anisotropy,\cite{Timm:2007} and higher order terms $(-B_{2n}/2) ({\hat S}_+^{2n} + {\hat S}_-^{2n})$ are neglected (they are usually small).

The second term corresponds to the superconducting parts, described by the BCS Hamiltonian 
\begin{equation}
	\mathcal{\hat H}_{\rm s} = 
	\sum\limits_{\ell, k, \sigma} \epsilon_k
	{\hat c}_{\ell, k, \sigma}^\dag {\hat c}_{\ell, k, \sigma}^{\phantom\dag}
	- \sum\limits_{\ell, k} \Delta_{\ell,k}
	\Bigl[ {\hat c}_{\ell, k, \uparrow}^\dag {\hat c}_{\ell, -k, \downarrow}^\dag + {\rm H.c.} \Bigr],
	\label{eq:sH}
\end{equation}
where $\epsilon_k = {\hbar}^2 k^2/2m - E_{\rm\scriptscriptstyle F}$ is the dispersion relation for free electrons, ${\hat c}_{\ell, k, \sigma}^\dag$ and ${\hat c}_{\ell, k, \sigma}^{\phantom\dag}$ are electronic creation and annihilation operators in the superconductors, $\ell$ enumerates left ($\ell = {\rm L}$) and right ($\ell = {\rm R}$) leads, $\Delta_{{\rm\scriptscriptstyle L(R)},k} = \Delta e^{\pm i\varphi/2}$, with $\Delta$ the superconducting gap and~$\varphi$ the superconducting phase difference along the junction.

The last term is the tunnel Hamiltonian between the leads and the molecule
\begin{equation}
	\mathcal{\hat H}_{\rm t} = 
	\sum\limits_{\ell, k, \sigma}
	\Bigl[ t_{\ell, k} {\hat d}_\sigma^\dag {\hat c}_{\ell, k, \sigma}^{\phantom\dag} + {\rm H.c.} \Bigr],
	\label{eq:tH}
\end{equation}
where $t_{\ell, k}$ are the tunneling amplitudes. By performing a gauge transformation for ${\tilde t}_{{\rm\scriptscriptstyle L},k} = t_{{\rm\scriptscriptstyle L},k} e^{i\varphi/4}$, ${\tilde t}_{{\rm\scriptscriptstyle R},k} = t_{{\rm\scriptscriptstyle R},k} e^{-i\varphi/4}$ and simultaneously for ${\tilde {\hat c}}_{{\rm\scriptscriptstyle L}, k, \sigma} = {\tilde {\hat c}}_{{\rm\scriptscriptstyle L}, k, \sigma} e^{-i\varphi/4}$, ${\tilde {\hat c}}_{{\rm\scriptscriptstyle R}, k, \sigma} = {\tilde {\hat c}}_{{\rm\scriptscriptstyle R}, k, \sigma} e^{i\varphi/4}$, one can ``move'' the dependence on $\varphi$ from $\Delta_{\ell,k}$ to $t_{\ell, k}$ and ${\hat c}_{\ell, k, \sigma}$ in Eqs.~(\ref{eq:sH}) and~(\ref{eq:tH}).\cite{Spivak:1991} We also perform a Bogoliubov transformation\cite{Gennes:1964} to diagonalize the BCS Hamiltonian, which takes the following form:
\begin{equation}
	\mathcal{\hat H}_{\rm s} = 
	\sum\limits_{\ell, k, \sigma} E_k
	{\hat\gamma}_{\ell, k, \sigma}^\dag {\hat\gamma}_{\ell, k, \sigma}^{\phantom\dag}
	\label{eq:sHbt}
\end{equation}
and the tunneling Hamiltonian reads
\begin{equation}
	\mathcal{\hat H}_{\rm t} = 
	\sum\limits_{\ell, k, \sigma}
	\Bigl[
		{\tilde t}_{\ell, k} {\hat d}_\sigma^\dag \big(
			u_k {\hat\gamma}_{\ell, k, \sigma}^{\phantom\dag} + 
			\sigma \, v_k {\hat\gamma}_{\ell, k, -\sigma}^{\dag} 
		\big) + 
		{\rm H.c.} 
	\Bigr],
	\label{eq:tHbt}
\end{equation}
where ${\hat\gamma}_{\ell, k, \sigma}^{\dag}$ and ${\hat\gamma}_{\ell, k, \sigma}$ are the quasiparticle creation and annihilation operators, $u_k = \sqrt{(1+\epsilon_k/E_k)/2}$ and $v_k = \sqrt{(1-\epsilon_k/E_k)/2}$ are the electron and hole coefficients, and $E_k = \sqrt{\epsilon_k^2 + \Delta^2}$ is the energy dispersion. In the following calculations we will consider for simplicity the case of symmetric contacts, thus $t_{{\rm\scriptscriptstyle L},k} = t_{{\rm\scriptscriptstyle R},k} = t_k$.

\subsection{Specific Hamiltonian for the $S=1$ case}

In the following, we will for simplicity restrict our calculations to the case of a molecular spin with $S=1$, which is the smallest value where easy axis anisotropy (parameter $D$) and QTM (parameter $B_2$) are nontrivial. As the electron occupation of the level is restricted to $0$ or $1$, let us write explicitly the molecule Hamiltonian in each case [see Eqs.~(\ref{eq:Hd}) and~(\ref{eq:Hm})].

For the empty electronic level, we have $\mathcal{H}_{\rm d}=\mathcal{H}_{{\rm m},0}$, and we use the basis $\{ |0\rangle_{\rm e} |1\rangle_{\rm m}, \; |0\rangle_{\rm e} |0\rangle_{\rm m}, \; |0\rangle_{\rm e} |-1\rangle_{\rm m} \}$, where $|0\rangle_{\rm e}$ represents the empty electronic state and $|S_z\rangle_{\rm m}$ the states of the molecule with spin projections $S_z = 1, 0, -1$. The matrix elements of $\mathcal{H}_{\rm m}$ are
\begin{equation}
	\mathcal{H}_{\rm m,0} = 
	\left[ \!
	\begin{array}{ccc}
		B-D & 0 & -B_2 \\
		0 & 0 & 0 \\
		-B_2 & 0 & -B-D \\
	\end{array}
	\! \right] \!.
	\label{eq:uH_me}
\end{equation}
The eigenvalues are noted $E_{0,i}$ ($i=1,2,3$), and the corresponding eigenvectors are ${\bf b}_i$. Below we will use the matrix $b_{ij} = [{\bf b}_1, \; {\bf b}_2, \; {\bf b}_3]$, which consists of columns of eigenvectors (first index enumerates columns, the second enumerates rows), and the inverse matrix ${\tilde b}_{ij} = (b_{ij})^{-1}$.

When the electronic level is occupied by one electron, we have $\mathcal{H}_{{\rm d},1} = \mathcal{H}_{{\rm m},1} + \epsilon_{\rm d}$, and we use the uncoupled spin basis ${ |s\rangle_{\rm e} |S_z\rangle_{\rm m} }$ (with $s=\uparrow,\downarrow$ and $S_z=+1,0,-1$). The matrix representation of $\mathcal{H}_{{\rm m},1}$ can then be decomposed as two independent $3\times 3$ submatrices:\cite{Huertas-Hernando:2006} $\mathcal{H}_{\rm m,1} = {\rm diag} \{\mathcal{H}_{\rm m,1}^+, \mathcal{H}_{\rm m,1}^- \}$, with
\begin{equation}
	\mathcal{H}_{\rm m,1}^+ = 
	\left[ \!
	\begin{array}{ccc}
		3B/2+J/2-D & 0 & -B_2 \\
		0 & -B/2 & J/\sqrt{2} \\
		-B_2 & J/\sqrt{2} & -B/2-J/2-D \\
	\end{array}
	\! \right]
	\label{eq:oHp_me}
\end{equation}
in the basis $\{ |\!\uparrow\rangle_{\rm e} |1\rangle_{\rm m}, \; |\!\downarrow\rangle_{\rm e} |0\rangle_{\rm m}, \; |\!\uparrow\rangle_{\rm e} |-1\rangle_{\rm m} \}$ and 
\begin{equation}
	\mathcal{H}_{\rm m,1}^- = 
	\left[ \!
	\begin{array}{ccc}
		B/2-J/2-D & J/\sqrt{2} & -B_2 \\
		J/\sqrt{2} & B/2 & 0 \\
		-B_2 & 0 & -3B/2+J/2-D \\
	\end{array}
	\! \right]
	\label{eq:oHm_me}
\end{equation}
in the basis $\{ |\!\downarrow\rangle_{\rm e} |1\rangle_{\rm m}, \; |\!\uparrow\rangle_{\rm e} |0\rangle_{\rm m}, \; |\!\downarrow\rangle_{\rm e} |-1\rangle_{\rm m} \}$. These matrices have eigenvalues $E_{1,i}^+$, $E_{1,i}^-$ and corresponding eigenvectors ${\bf a}_{i}^+$, ${\bf a}_{i}^-$. As previously, we define matrices $a_{ij}^{\pm} = [{\bf a}_1^{\pm}, \; {\bf a}_2^{\pm}, \; {\bf a}_3^{\pm}]$, and inverse matrices ${\tilde a}_{ij}^{\pm} = (a_{ij}^{\pm})^{-1}$.

\subsection{Josephson current}

The Josephson current through the molecule can be calculated using perturbation theory in the tunneling Hamiltonian $\mathcal{\hat H}_{\rm t}$;\cite{Spivak:1991} the first nonvanishing term is given by
\begin{multline}
	I = \frac{2e}{\hbar} \frac{\partial}{\partial\varphi}
	\bigl\langle {\rm gs} \big|
		\mathcal{\hat H}_{\rm t}
		(E_{\rm gs} - \mathcal{\hat H}_0)^{-1}
		\mathcal{\hat H}_{\rm t}
		(E_{\rm gs} - \mathcal{\hat H}_0)^{-1} \\
		\times \mathcal{\hat H}_{\rm t}
		(E_{\rm gs} - \mathcal{\hat H}_0)^{-1}
		\mathcal{\hat H}_{\rm t}
	\bigr|{\rm gs} \big\rangle,
	\label{eq:I_pert}
\end{multline}
where $\mathcal{\hat H}_0 = \mathcal{\hat H}_{\rm d} + \mathcal{\hat H}_{\rm s}$. The ground state $|{\rm gs}\rangle$ is the occupied state with lowest energy, thus it has energy $E_{\rm gs} = {\rm min} \{ E_{1,i}^\pm \}$, and $|{\rm gs}\rangle = |{\bf a}_i^\zeta\rangle$, where $i = 1, 2, 3$ specifies the state number and $\zeta=\pm$ is the block index. Note that the dot-lead coupling induces energy shifts for the occupied states of the dot, starting at order 2 in $\mathcal{\hat H}_{\rm t}$. However, we do not need to compute these shifts, as they will be identical for the two single occupied states, and they can be included in the value of $\epsilon_{\rm d}$ (see Ref.~\onlinecite{Lee:2010} for a multilevel case where these shifts have to be computed).

As was shown in Ref.~\onlinecite{Spivak:1991}, in the absence of coupling to a molecular spin, the perturbative approach allows us to understand the $\pi$ state due to large Coulomb interaction on the dot: the order of the electrons of a Cooper pair is necessarily reversed during tunneling through the dot, which gives opposite sign for the current due to the singlet nature of the Cooper pair. Here, the exchange coupling between the electron spin and a molecular spin means that the occupied state of the dot is a linear combination of states involving in general both $|\!\uparrow\rangle$ and $|\!\downarrow\rangle$ states of the electron spin. This creates the possibility of spin-flip processes: a spin-up electron tunneling in the dot can tunnel out as a spin-down electron for example. With such a spin flip, it is now possible for a Cooper pair to tunnel through the dot without reversing the order of electrons, thus contributing to positive current. In the presence of exchange coupling with a molecular spin, one can thus expect that, among all the lowest-order processes contributing to the Josephson current, some of them will contribute to negative current, and some others to positive current. The global sign of the current will thus depend on the relative weight of the different processes, which are a function of the parameters of the molecule Hamiltonian.
 
Expressing in Eq.~(\ref{eq:I_pert}) the action of the tunneling Hamiltonian on the eigenstates introduced in the previous section, a lengthy but straightforward calculation gives eventually
\begin{align}
	I = -\frac{4e}{\hbar} \sin\varphi
	\sum\limits_{k, k'} &
	t_{{\rm\scriptscriptstyle L}, k}^2 t_{{\rm\scriptscriptstyle R}, k'}^2 
	u_k v_k u_{k'} v_{k'} \nonumber \\
	\times \sum_j \biggl\{
	& \frac{
		A_{j,k'}^{\zeta*} B_{j,k}^\zeta +
		B_{j,k'}^{\zeta*} A_{j,k}^\zeta
	}{
		E_k + E_{k'} + E_{1,j}^\zeta - E_{\rm gs}
	} \nonumber \\
	+ & \frac{
		A_{j,k'}^{{\bar\zeta}*} A_{j,k}^{\bar\zeta} +
		B_{j,k'}^{{\bar\zeta}*} B_{j,k}^{\bar\zeta}
	}{
		E_k + E_{k'} + E_{1,j}^{\bar\zeta} - E_{\rm gs}
	}
	\biggr\}.
	\label{eq:I_fin}
\end{align}
Here
\begin{align}
	A_{j,k}^{\pm} = &
	\pm \frac{
		({\tilde a}_{{\rm gs},1} b_{11} + {\tilde a}_{{\rm gs},3} b_{31})
		({\tilde b}_{11} a_{1j}^\pm + {\tilde b}_{13} a_{3j}^\pm)
	}{
		E_k + E_{0,1} - E_{\rm gs} - \epsilon_{\rm d}
	} \nonumber \\
	& \pm \frac{
		({\tilde a}_{{\rm gs},1} b_{13} + {\tilde a}_{{\rm gs},3} b_{33})
		({\tilde b}_{31} a_{1j}^\pm + {\tilde b}_{33} a_{3j}^\pm)
	}{
		E_k + E_{0,3} - E_{\rm gs} - \epsilon_{\rm d}
	},
	\label{A} \\
	B_{j,k}^{\pm} = &
	\mp \frac{
		{\tilde a}_{{\rm gs},2} a_{2j}^\pm
	}{
		E_k + E_{0,2} - E_{\rm gs} - \epsilon_{\rm d}
	},
	\label{eq:B}
\end{align}
where $A_{j,k}^{\pm*} \equiv (A_{j,k}^\pm)^*$ and ${\bar\zeta} = -\zeta$. Equations~(\ref{eq:I_fin})--(\ref{eq:B}) represent the main results of this paper. Because we have performed a lowest-order tunneling calculation, we get a simple $I = I_{\rm c} \sin\varphi$ dependence of the current. However, the study of value of the critical current $I_{\rm c}$ (in addition to its sign) will give us precious information on the system. At zero temperature the sums over $k$ and $k'$ should be taken over the energy region $\epsilon_k, \epsilon_{k'} > 0$. Both summations over $k$ can be replaced by the integration over energy $\epsilon$: $\sum_k \to \int d\epsilon \rho(\epsilon)$, where $\rho(\epsilon)$ is a density of states.

Our formulas of course contain the known result for the case where there is no molecular spin and no magnetic field ($B = B_2 = J = D = 0$):\cite{Spivak:1991:comment} we obtain a $\pi$-junction with negative critical current,
\begin{align}
	I_{\rm c}^{(0)} 
	& = -\frac{4e}{\hbar} 
	\sum\limits_{k, k'}
	\frac{
		t_{{\rm\scriptscriptstyle L}, k}^2 t_{{\rm\scriptscriptstyle R}, k'}^2 
		u_k v_k u_{k'} v_{k'} 
	}{
		(E_k + E_{k'}) (E_k - \epsilon_{\rm d}) (E_{k'} - \epsilon_{\rm d})
	}
	\nonumber \\
	& \!\!\!\! = -\frac{4e}{\hbar} 
	\frac{\Gamma_{\rm\scriptscriptstyle L}\Gamma_{\rm\scriptscriptstyle R} \Delta^2}{4\pi^2} \!
	\int\limits_0^\infty \!
	\frac{d\epsilon_1 d\epsilon_2}{
		E_1 E_2 (E_1 + E_2) (E_1 - \epsilon_{\rm d}) (E_2 - \epsilon_{\rm d})
	},
	\label{eq:I_0}
\end{align}
where we assume constant density of states $\rho(\epsilon) = \rho_0 = 2m/\pi\hbar^2$, tunneling rates $\Gamma_{\rm\scriptscriptstyle L(R)} = \pi\rho_0 t_{\rm\scriptscriptstyle L(R)}^2$, and $E_{1(2)} = \sqrt{\Delta^2 + \epsilon_{1(2)}^2}$.

\begin{figure}[t]
	\includegraphics[width=\linewidth]{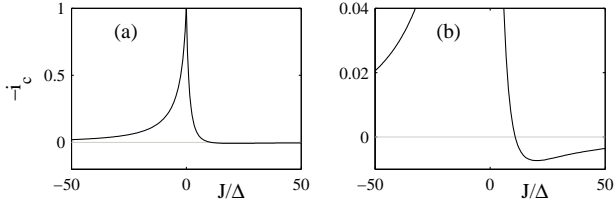}
	\vspace{-6mm}
	\caption{
The dependence of the normalized critical current~$i_{\rm c}$ as a function of the exchange coupling $J$ for an isotropic magnetic molecule ($D = B_2 = 0)$ and dot level $\epsilon_{\rm d}/\Delta = -5$ in the absence of magnetic field ($B = 0$).
	}
	\label{fig:I_J}
\end{figure}

In the next section we analyze the dependence of the dimensionless critical current $i_{\rm c} = I_{\rm c} / |I_{\rm c}^{(0)}|$ on the magnetic molecule parameters $J$, $D$, $B_2$, dot energy $\epsilon_{\rm d}$, and external magnetic field $B$. Positive $i_{\rm c}>0$ corresponds to the $0$-junction phase, negative $i_{\rm c}<0$ corresponds to the $\pi$-junction phase.

\section{Results and discussion}
\label{sec:results}

\begin{figure}[t]
	\includegraphics[width=\linewidth]{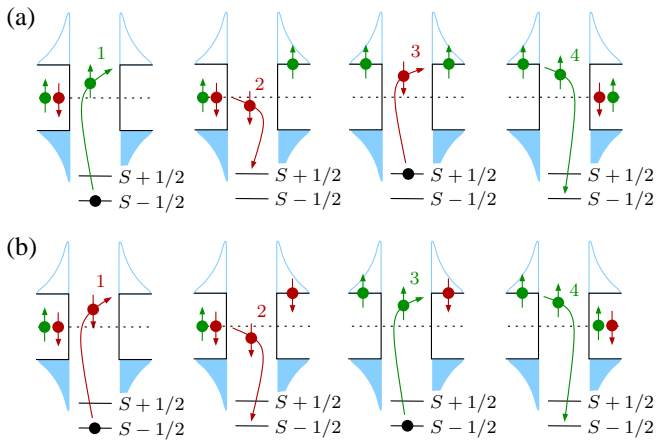}
	\caption{
(Color online) Illustration of two typical tunneling processes leading to the transfer of a Cooper pair. The presence of strong Coulomb interaction prohibits the double occupation on the dot and the electrons are transferred one by one. Because of the exchange coupling $J$ between the electron spin and the molecular spin, the state of the occupied dot (black circle on the figure) is characterized by the total spin, $S-1/2$ and $S+1/2$. These two levels are separated by an energy $3J/2$. The process where the intermediate state of the occupied dot (a) is different from the initial one and the process where the intermediate state is the same as the initial one (b) contribute with different signs to the Josephson current. The competition between these two processes leads to the existence of the $\pi$--$0$ transition; see Eq.~(\ref{eq:I_J}).
	}
	\label{fig:spin_flips}
\end{figure}

For reference, we start by analyzing Eq.~(\ref{eq:I_fin}) as a function of exchange coupling $J$, when no anisotropy is present ($D = B_2 = 0$) and without magnetic field ($B=0$). As shown in Fig.~\ref{fig:I_J}(a), the current is suppressed both by negative and positive $J$. For negative~$J$ (ferromagnetic coupling) the system always remains in the $\pi$ state ($i_{\rm c} < 0$). For positive $J$ (antiferromagnetic coupling) a $\pi$--$0$ transition occurs for $J/\Delta \sim 10$ (the precise value is slowly varying with $\epsilon_{\rm d}$). This behavior can be understood by looking at the formula for the current,\cite{Lee:2008a}
\begin{align}
	I = -\frac{4e}{\hbar} \sin\varphi
	\sum\limits_{k, k'} &
	t_{{\rm\scriptscriptstyle L}, k}^2 t_{{\rm\scriptscriptstyle R}, k'}^2 
	u_k v_k u_{k'} v_{k'} \nonumber \\
	\times \frac{1}{3 \mathcal{E}_k \mathcal{E}_{k'}} \biggl\{
	& \frac{4}{3J/2+E_k+E_{k'}} -
	\frac{1}{E_k+E_{k'}}
	\biggr\},
	\label{eq:I_J}
\end{align}
where $\mathcal{E}_k = E_k + J - \epsilon_{\rm d}$. The first term depicts the transfer of a Cooper pair involving a change of the total coupled spin (electronic and molecule) during the intermediate state [e.g., see Fig.~\ref{fig:spin_flips}(a)], while the second term corresponds to a Cooper pair without change of total spin during the intermediate state [e.g., see Fig.~\ref{fig:spin_flips}(b)]. For large positive $J$, the first term becomes smaller than the second one, and the sign of the current changes, which explains the $\pi$--$0$ transition.

Note that there is no change of ground state associated with this transition occurring for large positive $J$, hence the critical current shows a smooth change from negative to positive value, passing continuously through arbitrary small values. This is to be contrasted with $0$--$\pi$ transition, which is due to the crossing of energy levels leading to a change of ground state,\cite{Rozhkov:2000,Novotny:2005,Sadovskyy:2007} where an abrupt change of the critical current can be observed (see, e.g., Figs.~4 and~6 in Ref.~\onlinecite{Lee:2010}).

\begin{figure}[t]
	\includegraphics[width=0.95\linewidth]{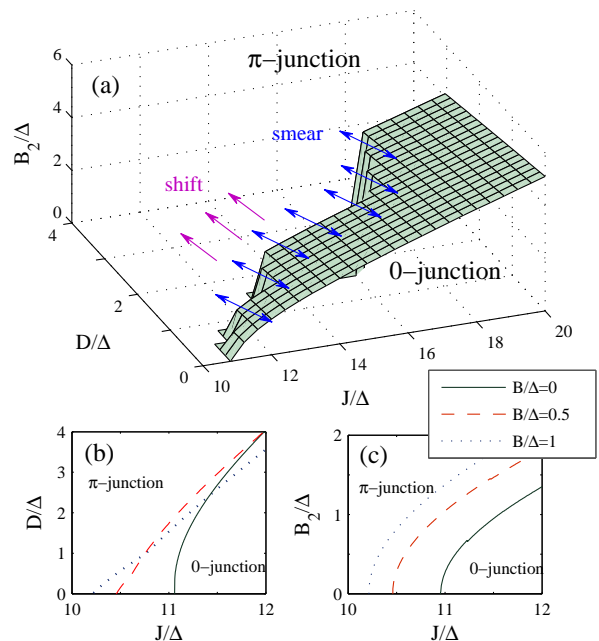}
	\vspace{-6mm}
	\caption{
(Color online) (a)~$\pi$- and $0$-junction regions as a function of the $J$, $D$, and $B_2$. The surface divides 3D space $(J, D, B_2)$ to the top region in a $\pi$-junction regime and to the bottom one with $0$-junction regime; at the surface current is zero. Magnetic field is zero, $B=0$, its increasing leads to ``shift'' and ``smear'' of the surface as shown by magenta and blue arrows.
(b)~$\pi$--$0$ transition diagram in $(J, D)$ space at $B_2 = 0$. Different curves correspond to the different $B$'s: $B/\Delta = 0.0$ (solid), $0.5$ (dashed), and $1.0$ (dotted).
(c)~$\pi$--0 transition diagram in $(J, B_2)$ space at different $B$ and $D = 0$. The nonzero $D$ and $B_2$ increases the critical $J_{\rm c}$ (see Fig.~\ref{fig:I_J}) and magnetic field mainly decreases $J_{\rm c}$.
	}
	\label{fig:J_D_B2}
\end{figure}

We will now consider the effect of the anisotropy ($D$ and $B_2$) and of the magnetic field~$B$ on the critical current, especially near the $\pi$--$0$ transition. We assume that the superconducting gap is independent of the magnetic field. Figure~\ref{fig:J_D_B2}(a) shows the effect of $D$ and $B_2$ on the transition; the surface shows the values of the parameter for which the current is zero. Above the surface the system is in the $\pi$-junction phase ($i_{\rm c}<0$), while under the surface the system is in the zero phase ($i_{\rm c}>0$). One can see that both $D$ and $B_2$ move the $\pi$--$0$ transition to higher values of $J$. This is confirmed by Figs.~\ref{fig:J_D_B2}(b) and~\ref{fig:J_D_B2}(c), which correspond to cuts of the three-dimensional (3D) plot for fixed values of $B_2$ and $D$, respectively. On these panels, the different curves correspond to different values of the magnetic field $B$: we see that increasing the magnetic field tends to push the system toward the 0-junction phase (note that the results are insensitive to the sign of~$B$). On the 3D plot Fig.~\ref{fig:J_D_B2}(a) the effect of the magnetic field $B$ is thus to shift the zero current surface as shown with magenta arrows, and also to somewhat smear the sharp behavior in $B_2$ as shown with blue arrows.

Up to now, we have studied the phase diagram of the system as a function of the exchange coupling $J$ and of the anisotropy parameters $D$ and $B_2$. However, for a given molecule, these parameters have usually a fixed value. We will now study the behavior of the critical current when the experimentally adjustable quantities, the external magnetic field $B$ and the dot level $\epsilon_{\rm d}$, are varied. The goal is to understand how the values of the exchange coupling and of the anisotropy parameters will modify the behavior of the current as a function of~$B$ and~$\epsilon_{\rm d}$. This could be an original way to obtain information on the exchange coupling and on the spin anisotropy in the molecule, by measuring the critical current of the tunnel junction and varying $B$ and $\epsilon_{\rm d}$.

\begin{figure*}[t]
	\includegraphics[width=0.80\textwidth]{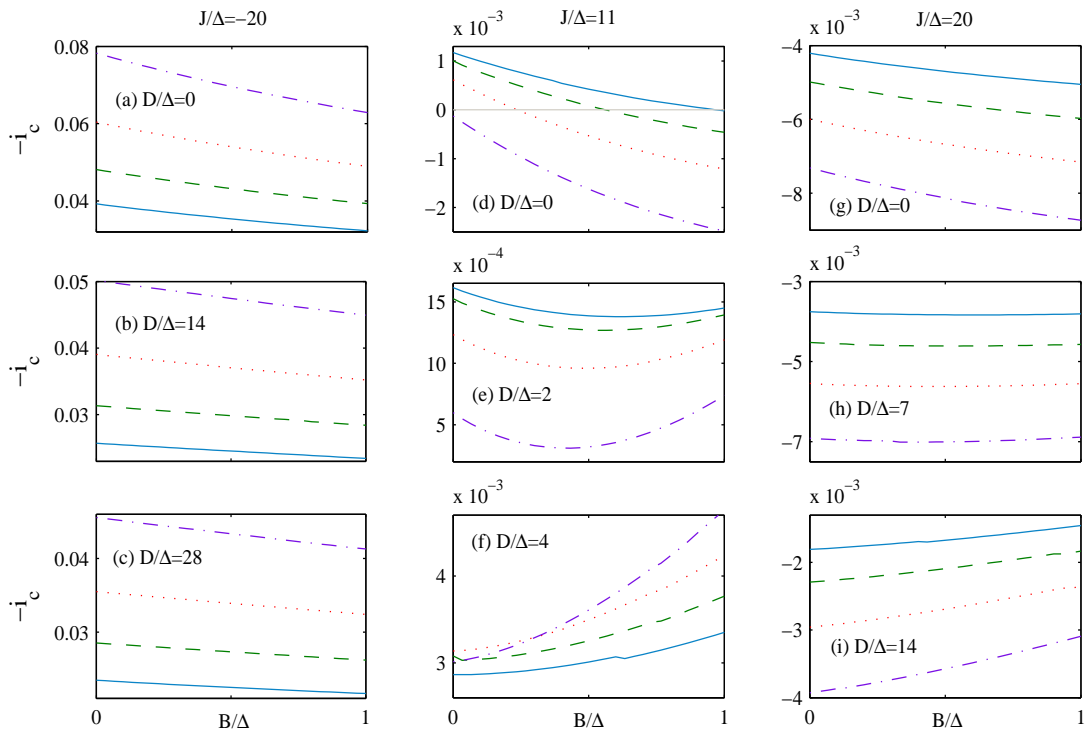}
	\caption{
(Color online) Critical current as a function of the magnetic field $B$ for different values of the exchange coupling $J$ and the anisotropy parameter $D$ ($B_2 = 0$). The different curves in a single plot are for various dot levels $\epsilon_{\rm d}$: $\epsilon_{\rm d}/\Delta = -12.5$ (solid cyan line), $-10.0$ (dashed green line), $-7.5$ (dotted red line), and $-5$ (dashed-dotted magenta line); all currents are normalized by the critical current obtained for $\epsilon_{\rm d}/\Delta = -5$ with $B = J = D = B_2 = 0$ [see Eq.~(\ref{eq:I_0})].
Left column: $J/\Delta = -20$; the system is deep in the $\pi$-junction regime, and the anisotropy parameter $D$ does not change the curves qualitatively.
Middle column: $J/\Delta = 11$; the system is near $\pi$--$0$ transition, and the value of $D$ has a great impact on the behavior of the curves: it can produce nonmonotonic behavior as a function of~$B$ [panel (e)], or reverse the slope of the curves compared to $D = 0$ [panel (f)].
Right column: $J/\Delta = 20$; the system is in the $0$-junction regime and $D$ still has a visible impact, as it can change the slopes of the curves [panel (h)]. This is due to the presence of a (large) critical value $D_{\rm c}$ above which the system is again in the $\pi$-junction phase (not shown).
	}
	\label{fig:I_B}
\end{figure*}

The different panels of Fig.~\ref{fig:I_B} show the behavior of the critical current as a function of~$B$ and for various values of the dot level $\epsilon_{\rm d}$, the exchange coupling $J$ and of the anisotropy parameter $D$ (for simplicity, we have taken $B_2 = 0$). Each column is for a given value of $J$: deep in the $\pi$-junction regime $J/\Delta = -20$ (left), in the intermediate regime $J/\Delta = 11$ (middle), and deep in the $0$-junction regime $J/\Delta = 20$ (right). The top panel of each column is for $D = 0$, while the two bottom panels of each column are for nonzero values of $D$ as indicated. The richer behavior is obtained when the exchange coupling has a value that allows us to observe the $\pi$--$0$ transition, here in the second column for $J/\Delta = 11$. Without anisotropy [Fig.~\ref{fig:I_B}(d)], we see that by sweeping the magnetic field we can observe the $\pi$--$0$ transition. In the presence of small anisotropy [Fig.~\ref{fig:I_B}(e)], we observe a nonmonotonic behavior as a function of~$B$, with the modulus of the critical current $|i_{\rm c}|$ decreasing as a function of~$B$ for small $B$, but increasing for large $B$. Finally, for larger anisotropy [Fig.~\ref{fig:I_B}(f)], $|i_{\rm c}|$ is everywhere increasing as a function of~$B$. Note that, between panel Fig.~\ref{fig:I_B}(d) ($D = 0$) and panel Fig.~\ref{fig:I_B}(f) ($D/\Delta = 4$), the order of the curves as a function of $\epsilon_{\rm d}$ has been reversed. When $J$ is much larger than the superconducting gap (right column, with $J/\Delta = 20$), the system is deep in the $0$-junction phase, but the anisotropy has a visible impact on the curves: comparing Fig.~\ref{fig:I_B}(g) (for $D/\Delta = 0$) with Figs.~\ref{fig:I_B}(h) and~\ref{fig:I_B}(i) (for $D/\Delta = 7$ and $14$), we see that when $D$ is large enough, the slope of the critical current is the opposite of the one for small~$D$. This is a consequence of the $\pi$--$0$ transition, which happens for larger $D$. Finally, for negative $J$ [Figs.~\ref{fig:I_B}(a)--\ref{fig:I_B}(c) with $J/\Delta = -20$], the anisotropy does not bring any qualitative change to the behavior of the current as a function of the magnetic field, and $|i_{\rm c}|$ always decrease with $B$.

From the different curves shown in Fig.~\ref{fig:I_B}, we can deduce that when $J$ is positive (the antiferromagnetic coupling case), the anisotropy has a visible impact on the behavior of the critical current as a function of~$B$, as it can produce a nonmonotonic behavior close to the $\pi$--$0$ transition, and reverse the slope of $|i_{\rm c}|$ as a function of~$B$ when $J$ is much larger than the critical value. On the other hand, for negative $J$ (ferromagnetic coupling), the anisotropy does not have a qualitative effect on the critical current, and it merely reduces the value of $|i_{\rm c}|$.

\section{Conclusion}
\label{sec:conclusion}

We have computed the Josephson current through a magnetic molecule in the tunneling regime, studying the effect of the exchange coupling with the molecular spin, and the spin anisotropy of the molecule. Performing a perturbative calculation starting from a Hamiltonian model, we have shown that an antiferromagnetic coupling between the electron spin and the molecular spin can induce a $\pi$--$0$ transition. We have described how the spin anisotropy $D$ and the quantum tunneling of magnetization term $B_2$ affect the transition.

We have shown that by studying the behavior of the critical current as a function of the magnetic field and the level position (which are both experimentally tunable parameters), it is possible to get information on the value of the spin anisotropy $D$, even outside the range of the $\pi$--$0$ transition.

This work could be extended in several directions. The calculations could be performed for a larger molecular spin (albeit at the cost of heavier expressions). One could also use anisotropy parameters that depend on the charge state of the molecule (and thus on the occupation of the dot in our model), which could describe more faithfully molecular magnets like Mn$_{12}$.\cite{Heersche:2006b} One could also consider the case of an external magnetic field aligned along an arbitrary direction (and not along the anisotropy axis of the molecule), in order to describe experiments where it is not possible to control the anisotropy orientation. Such a magnetic field should have a strong impact on the current, as it will mix efficiently all the molecular states.\cite{Timm:2007}

Finally, new possibilities could open up if one considers explicitly the Josephson current between type II superconductors. In this case, it could be possible to control the value of the superconducting gap $\Delta$ with the applied magnetic field. Going to very small $\Delta$ would give large values of $J/\Delta$, $D/\Delta$, etc., and a very large parameter range of the system, including the $\pi$--$0$ transition for $J>0$, could be explored. In the same manner, it is possible to enhance the critical temperature $T_{\rm c}$ and the second critical field $H_{\rm c2}$ by decreasing the thickness of the superconductor.\cite{Meservey:1971} This could allow us to use large values of the magnetic field.

We acknowledge financial support by the CNRS LIA agreements with Landau Institute; Grants No.~NSF ECS-0608842, No.~ARO W911NF-09-1-0395, No.~DARPA HR0011-09-1-0009 (I.A.S.); NRF 2011-0003495 (M.L.); and JST-CREST, the ``Topological Quantum Phenomena'' (No.~22103002) KAKENHI on Innovative Areas, and a Grant-in-Aid for Scientific Research (No.~22710096) from MEXT of Japan (S.K.).


\begin{thebibliography}{99}
\bibitem{Spivak:1991:comment}
	Note that result Eq.~(\ref{eq:I_0}) disagrees 
	with Ref.~\onlinecite{Spivak:1991}, where a similar 
	calculation was done; in the answer of this paper 
	the expression $u_k^2 v_{k'}^2 + u_{k'}^2 v_{k}^2$ 
	should be replaced by $u_k v_k u_{k'} v_{k'}$.
\end{thebibliography}
\end{document}